\documentclass[10pt,osajnl2,twocolumn,showpacs,superscriptaddress]{revtex4} 

\usepackage{amsmath}
\usepackage{amssymb}
\usepackage{graphics}
\usepackage{hyperref}

\begin{document}

\title{Extended cavity diode lasers with tracked resonances}

\author{Sheng-wey Chiow}
\email{swchiow@stanford.edu}
\author{Quan Long}
\author{Christoph Vo}
\author{Holger M\"uller}
\affiliation{Physics Department, Stanford University, Stanford, CA 94305, USA}
\author{Steven Chu}
\affiliation{Physics Department, Stanford University, Stanford, CA 94305, USA} \affiliation{Lawrence Berkeley National Laboratory and Department of Physics,
University of California, Berkeley, Berkeley, CA 94720, USA}

\begin{abstract}
We present a painless, almost-free upgrade to present extended cavity diode lasers (ECDLs), which improves the long term mode-hop free performance by
stabilizing the resonance of the internal cavity to the external cavity. This stabilization is based on the observation that the frequency or amplitude noise
of the ECDL is lowest at the optimum laser diode temperature or injection current. Thus, keeping the diode current at the level where the noise is lowest
ensures mode-hop free operation within one of the stable regions of the mode chart, even if these should drift due to external influences. This method can be
applied directly to existing laser systems without modifying the optical setup. We demonstrate the method in two ECDLs stabilized to vapor cells at 852\,nm and
895\,nm wavelength. We achieve long term mode-hop free operation and low noise at low power consumption, even with an inexpensive non-antireflection coated
diode.
\end{abstract}


\maketitle



Diode lasers are compact, have low power consumption, are readily available in certain wavelength ranges and are relatively cheap. Their free running output
frequency $\nu_L$ is defined by the resonance frequency $\nu_i$ of the standing waves within the diode. However, this depends on the temperature $T_d$ of the
diode, the injection current $I_d$, and other influences such as aging. For many applications like in atomic physics~\cite{Wiemann}, the stability of $\nu_L$
needs to be improved. In extended cavity diode lasers (ECDLs), this is achieved by retroreflection of a part of the output beam from a diffraction
grating~\cite{Wiemann,Preston81,Labachelerie,Arnold,Ricci}. $\nu_L$ is now determined primarily by one of the resonance frequencies $\nu_e^m$ (where $m$ is a
mode number) of the external cavity formed by the diode's back facet and the grating. The particular mode $m$ of the external cavity is selected by $\nu_i$,
the frequency defined by the reflection angle of the grating $\nu_g$, and other (parasitic) resonances such as between the diode output facet and the diode
housing's window. For stable operation, ideally all resonances are overlapped; in particular, $\nu_e^m\simeq\nu_g\simeq\nu_i$. The frequency stability of ECDLs
can furthermore be increased by stabilizing (``locking'') it to atomic or molecular resonances.

While such locking reduces the drift of $\nu_L$, external influences on the various resonance frequencies can offset their overlap, eventually causing
mode-hops or broadband multimode operation. This offset of overlap cannot be compensated with conventional frequency locking scheme, since only the output frequency is controlled, not the relative alignment of $\nu_e^m, \nu_g\,$ and $\nu_i$. The following steps are often taken to enable mode-hop free operation: Antireflection (AR)-coating of the laser
diode's output facet reduces the finesse of the internal cavity, so that the stable regions for $\nu_i$ become larger~\cite{Zorabedian,Sacher}. Removing the
window of the diode case reduces the number of parasitic cavities. Moreover, extensive shielding from environmental influences is necessary, which includes a
hermetically sealed housing of the whole setup and stabilization of the housing temperature as well as $T_d$ by two thermostats. Allard {\em et al.} use a
computer-controlled setup~\cite{Allard} that can automatically reset the laser after a mode-hop.

However, AR-coated diodes are one or two orders of magnitude more expensive then non-coated ones, and may not be available at all wavelengths. Removal of the
housing window opens the hermetic protection of the diode, thus affecting the reliability. The energy consumption and heat dissipation of the housing
temperature controller strongly exceeds the one of the diode itself. These may be obstacles if particular wavelengths, high reliability, or battery-powered
operation is required.

Passive feed-forward, e.g., choosing an appropriate pivot point
for rotating the grating~\cite{Favre,Labachelerie2}, has been used
to adjust $\nu_g$ in order to increase the mode-hop free tuning
range in ECDLs. Multiple actuators on the
grating~\cite{Lonsdale,Petridis}, current
feed-forward~\cite{Hult}, or electro-optic prisms~\cite{Menager}
have been used. While these methods can account for the
predictable part of the relative variations, they do not avoid
mode-hops due to environmental influences.

An alternative approach to facilitate single-mode operation is the use of very short external cavities, in particular by using the back emission from the rear
facets for the optical feedback. These have a large free spectral range and thus the overlap of $\nu_i$ and $\nu_e^m$ can drift more before a mode-hop occurs.
Besides this purely passive approach, active stabilization of the overlap of $\nu_i$ and $\nu_e^m$ has been used. For example, one can stabilize the laser
power $P$ via feedback to $I_d$, and stabilizing the mode by monitoring $dP/dI_d$ and feeding back to $T_d$~\cite{Liou85}. $\nu_e^m$ can also be stabilized by
optimizing $P$, which is inferred from the laser diode's voltage drop~\cite{Cassidy88}. Also, the pattern formed by the interference between the front and back
emissions of a diode laser has been used to monitor the mode and to stabilize the laser~\cite{Ventrudo93}. However, short-external-cavity diode lasers have an
increased linewidth, because of the lower selectivity of the shortened cavity. Also, the light from the back facet is not commonly accessible in diode laser
systems, especially those used in atomic physics where stability is crucial. 


\begin{figure}
\centerline{\resizebox{0.48\textwidth}{!}{%
  \includegraphics{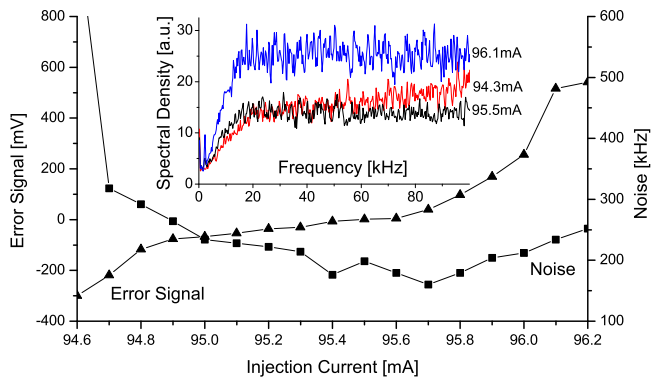}}}
\caption{\label{noisevscurrent} RMS frequency noise (integrated
from 40-100\,kHz) 
and the corresponding error signal versus injection current. Inset: Noise spectra for three injection
current settings.}
\end{figure}

Here, we present a new method for achieving stable mode-hop free operation in ECDLs: It is based on tracking the internal cavity's resonance at $\nu_i$
according to the external cavity at $\nu_e^m$ to keep the resonance frequencies overlapped. Moreover, this method keeps the ECDLs operated at the lowest noise
point in a mode-hop free band. (We note that modulation techniques have also been used to overcome systematic effects in frequency
standards~\cite{OCAMS,MuellerAPB}.) Furthermore, it can be adopted to existing laser systems without modifying the optical setup.

The method is based on our observation that the frequency and amplitude noise of the ECDL increases strongly near a mode-hop. For example, Fig.
\ref{noisevscurrent} shows the rms frequency noise as measured against a hyperfine line of Cs in a vapor cell using modulation transfer spectroscopy (MTS) (ac-coupled feedback to the injection current and dc-coupled to the grating via a piezo),
integrated from 40-100\,kHz, versus injection current. The injection current range of 94.6-96.2\,mA corresponds to one full band of mode-hop free operation.
It is obvious that the noise is increasing at the ends of the band and has a minimum in between. It is desirable to operate the laser at the point of lowest
noise. As explained, the stable band drifts due to changing environmental influences. Thus, the current has to be adjusted to keep at the desired point of
operation.

\begin{figure}
\centerline{\resizebox{0.48\textwidth}{!}{%
  \includegraphics{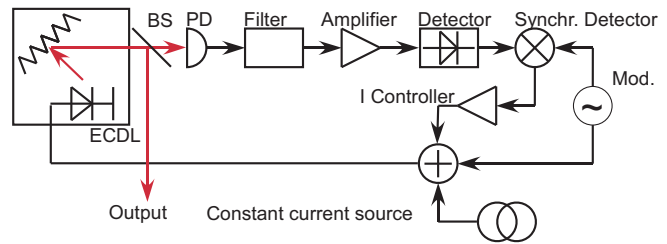}}}
\caption{\label{scheme} Schematic of a tracked ECDL. BS,
beamsplitter.}
\end{figure}

Therefore, a tiny modulation is added to $I_d$ (Fig. \ref{scheme}); it is small and slow enough ($\sim10$\,Hz) so that feedback to the piezo will remove the
frequency modulation (FM) caused by this in a frequency-locked system. Thus, $\nu_L$ is still stabilized to an atomic transition while the noise is probed at
different points in one mode-hop free region. The noise of the laser is measured by a photodetector PD, band-pass filtered for detecting the spectral range of
the noise showing the most pronounced effect, amplified, and its amplitude detected by a synchronous detector which measures the correlation of the amplitude
and the modulation. This results in an error signal which is zero when the noise is lowest (Fig. \ref{noisevscurrent}). An integral (I-) controller ensures
$\nu_i=\nu_e^m$ by controlling $I_d$ and/or $T_d$ (a ``tracked'' ECDL). This prevents the relative drift of the resonances in the first place, so that mode-hop
free operation is facilitated.

Note that the amplitude noise exhibits a similar behavior as the frequency noise shown in Fig. \ref{noisevscurrent}. Therefore, the amplitude noise can be used
instead of the frequency noise. In applications of frequency stabilization, like the ones considered here, a signal for the frequency noise may be available
anyway. In most other applications, however, it will be much easier to use the amplitude noise instead. In the following, we will describe one example for
each: Both lasers are locked to a Cs vapor cell using MTS, and one uses a cheap, non-AR-coated laser diode that is not equipped with a dual thermostat.

\begin{figure}
\centerline{\resizebox{0.48\textwidth}{!}{%
  \includegraphics{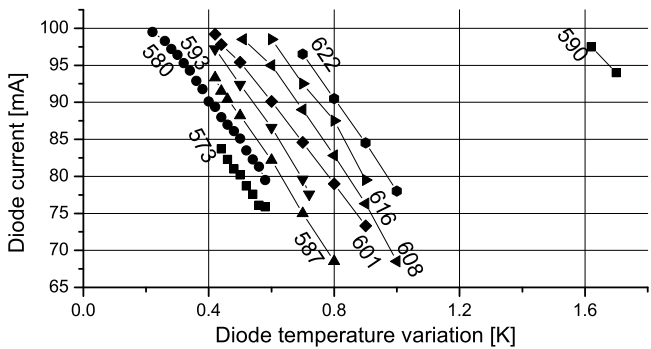}}}
\caption{\label{modechart} Mode chart of an ECDL for fixed
grating. Each graph represents a mode of the external cavity,
labelled with the wavelength $\lambda-894$\,nm at the center in pm ($\lambda$
changes by about 6\,pm in each mode). The 7\,pm mode spacing (3\,GHz) matches well with the free spectral range of the external cavity length of 5.5\,cm. The
graph on the far right hand side corresponds to a different mode
of the internal cavity([17\,pm free spectral range).}
\end{figure}

For the first tracked ECDL, we use a L904P030 type laser diode
sold by Thorlabs, Inc. It has an index-guided multi-quantum well
structure~\cite{Burkard}. The specifications include a wavelength
range of 885-920nm, a threshold injection current of 40\,mA and a
maximum output power of 30\,mW. We purchased 10 diodes and
selected one whose intrinsic wavelength was close to 895\,nm. The
diode (with the housing window not removed) and its collimating
lens are mounted on a thermoelectrical cooler for temperature
stabilization. We use a gold coated grating having
18,000\,lines/cm at a distance of about 5.5\,cm. The grating is
mounted on a Lees mirror mount attached to an invar base plate. An
additional invar rod connects the mirror mount and the diode
holder on their tops. The setup is protected from air currents and
acoustics by a hermetically sealed aluminum can with a Brewster
window. Fig. \ref{modechart} shows stable modes as a function of
$T_d$ and $I_d$.

The laser is locked to a hyperfine line of the D1 manifold in a Cs vapor cell by MTS using the usual methods~\cite{Shirley}. The FM at 80\,kHz used for the MTS
is generated by modulating the frequency of an acousto-optic modulator in the double-pass configuration. The feedback to the laser frequency is acting on $I_d$
for small, fast corrections (with a sensitivity of about 300\,MHz/mA), and to a piezo attached to the grating for slow corrections ($\lesssim 10$\,Hz), with a
sensitivity of about 2\,GHz/V. $I_d$ is controlled by the circuit described in~\cite{Libbrecht}. However, the influences of the environment are especially
strong for an ECDL using an uncoated diode; mode-hop free operation over more than about 1/2 h is not achieved.

We thus use tracking of $\nu_i$ for continuous stable operation
(Fig. \ref{scheme}). We modulate $I_d$ by $0.01-0.1$\,mA peak at
$\sim 10\,$Hz around its mean value near 100\,mA. A photodetector
detects the amplitude noise in the frequency range up to 25\,MHz
(the same photodetector is also used to generate the MTS signal).
Its output signal is high-pass filtered to reject the MTS
modulation frequency, amplified by two cascaded ZFL-500LN
amplifiers (Mini-Circuits) and its amplitude detected using a
Schottky diode peak detector. The rectified signal thus generated
is applied to a commercial lock-in amplifier to detect the
correlation to the current modulation. To control the internal
cavity we use an integral controller having a time constant of
about 10\,s. $T_d$ and $I_d$ are changed simultaneously so that
the ECDL keeps operating within a stable region of Fig.
\ref{modechart} even for large corrections. The low current
modulation frequency enables the Cs lock to take out any FM via
adjusting the grating.

\begin{figure}
\centerline{\resizebox{0.48\textwidth}{!}{%
  \includegraphics{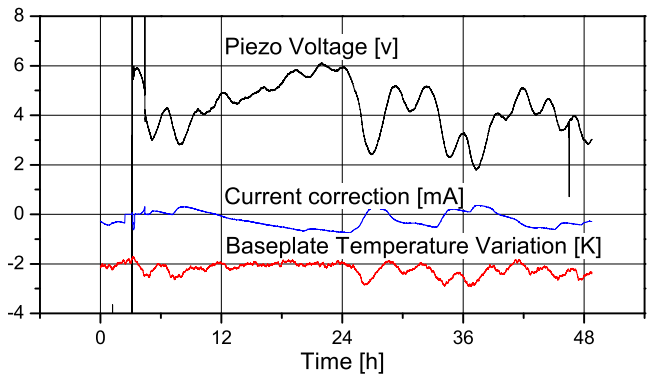}}}
\caption{\label{lock48h} 48-h operation of the non-AR-coated
laser.}
\end{figure}

This method is able to avoid the mode-hops associated with the non-AR-coated diode over long periods of time: Fig. \ref{lock48h} shows time traces from 48\,h
of operation locked to the Cs vapor cell. The piezo voltage indicates the operation of the Cs frequency lock. Also shown are the variations of the laboratory
temperature, measured at the base plate of the laser, and the correction applied to $I_d$. As can be seen, all parameters are partially correlated. However,
the correlation is not perfect (due to other environmental influences), so simple feed-forward to $I_d$ would not be able to avoid mode-hops in all cases.

\begin{figure}
\centerline{\resizebox{0.48\textwidth}{!}{%
  \includegraphics{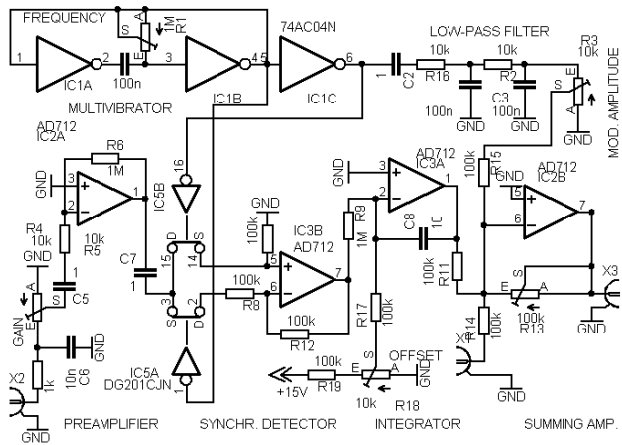}}}
\caption{\label{circuit} Lock-in amplifier and servo circuit.}
\end{figure}

One of the virtues of tracked ECDLs is the possibility of cheap, compact construction with low power dissipation. To take full advantage of these, we developed
the circuit shown in Fig. \ref{circuit}. IC1A and B form a multivibrator that generates a $\sim 10$\,Hz modulating frequency. This signal is low-pass filtered
to smooth the waveform and modulates $I_d$ via the current controller, which is connected to X3. The rectified diode laser noise is connected to X2. It is
low-pass filtered, ac coupled and then preamplified by IC2A. Synchronous detection is performed by IC3B and IC5: IC5A,B subsequently connect the signal to
either the positive or the negative input of IC3B. The output signal thus detected is applied to the integral controller formed by IC3A. A small offset set by
R18 can be applied, which is set such that the servo adjusts the diode current for optimum operation in the middle of the mode-hop free range. The summing
amplifier formed by IC2B adds the modulating signal, the servo output, and any external modulation (e.g., for use for the Cs lock).

We also use tracking for an ECDL that uses an AR-coated GaAlAs
diode (Type SDL-5411-G1), but no dual temperature stabilization.
The output power of the ECDL is $\sim 70$\,mW. The ECDL is
stabilized to a Cs vapor cell using MTS as described above, but
with a modulation frequency of 10\,MHz generated by an
electro-optical modulator. For the tracking, we minimize the
residual frequency noise as measured at the input of the servo
used for the MTS stabilization. As seen in Fig.
\ref{noisevscurrent} (inset), the noise spectrum below about
20\,kHz exhibits a monotonous increase with the injection current.
This reduces the increase of the rms noise also shown in Fig.
\ref{noisevscurrent} at the low current end of the mode-hop free
band. Therefore, these low-frequency parts of the noise spectrum
are suppressed by a high-pass filter to increase the slope of the
error signal before detecting its amplitude. The signal thus
generated is fed to the input of the circuit shown in Fig.
\ref{circuit}, which in turn applies small corrections to the
diode current to keep the noise lowest.

\begin{figure}
\centerline{\resizebox{0.48\textwidth}{!}{%
  \includegraphics{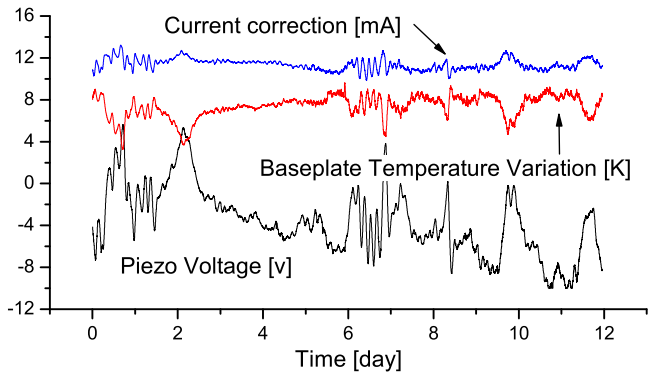}}}
\caption{\label{12daylock} 12-day operation of the AR-coated
laser.}
\end{figure}

Fig. \ref{12daylock} shows the action of the system. Without baseplate temperature stabilization, baseplate temperature changes would cause mode-hops every few
h of operation. However, tracking $\nu_i$ via the diode current enables 12\,days of uninterrupted, mode-hop free operation (after which the laser was
deliberately switched off). Also shown is the correction applied to the piezo by the MTS stabilization. Tracking also helps to keep the noise of the laser low.
Again, the low modulation frequency ensures that the Cs lock can take out any FM.

Note that this method is not restricted to vapor-cell stabilized diode laser systems as demonstrated above. In fact, it should be applicable to any diode laser
stabilized in any form with at least one feedback mechanism in addition to current feedback. This secondary mechanism is necessary to remove the FM otherwise
caused by our method, so that $\nu_L$ is still determined by the original stabilization machinery. For example, a frequency-/phase-locked diode laser with
current and piezo feedback can benefit from this method. Moreover, a diode laser stabilized to an ultra-stable optical cavity can adopt this method to acquire
better long-term stability. Even with an injection-locked diode laser, one can, in principle, modulate the passively regulated $I_d$ and correlate the
amplitude/phase noise for tracking the internal cavity.

To sum up, tracked ECDLs can be inexpensive, reliable and have low power consumption as they can be operated without a thermostat for the housing. They have
low frequency noise since optimum overlap of the external and internal resonances is maintained. Non-AR-coated diodes can be used with the case unopened. We
achieve mode-hop free low noise operation of tracked ECDLs while they are locked to Cs vapor cells. In daily laboratory use, we do not observe any mode-hops
(other than caused by external mechanical bumps). Moreover, our method is a painless, almost-free upgrade of current systems that can improve the time of
mode-hop free operation by at least a factor of 100 (from $1/2$\,h to $>48$\,h and from few h to $>12$\,days): All one needs is some more servo electronics.
Thus, we believe that it might find widespread applications. For the ultimate in performance, one can combine the method presented here with a state-of-the-art
mechanical design~\cite{Wiemann,Labachelerie,Arnold,Ricci} that is more compact and rigid. Another promising application of the method would be to increase the
mode-hop free tuning range of ECDLs.

\section*{Acknowledgments}
We thank Kurt Gibble who initially built the ECDL setups we used. This work is sponsored in part by grants from the AFOSR, the NSF and the MURI. H.M. wishes to
thank the Alexander von Humboldt-Stiftung for their valuable support.

\end{document}